\documentclass{appolb}
\usepackage{amsmath,amssymb,booktabs,multirow,graphicx,xcolor,float}
\usepackage[colorlinks=true,linkcolor=blue!50!black,citecolor=blue!50!black,urlcolor=blue!50!black]{hyperref}
\usepackage{placeins}
\usepackage[switch]{lineno}
\newcommand{\abinv}{\mathrm{ab}^{-1}}
\newcommand{\MET}{E_T^{\rm miss}}
\newcommand{\gone}[1]{g^{(1)}_{#1}}
\newcommand{\gtwo}[1]{g^{(2)}_{#1}}
\newcommand{\gt}[1]{\tilde g_{#1}}

\definecolor{actagreen}{RGB}{75,105,25}
\makeatletter
\def\ps@titpage{%
  \def\@oddfoot{\hfil\small(\ThePage)\hfil}\let\@evenfoot\@oddfoot
  \def\@oddhead{{\color{actagreen}\vbox{\hbox to\textwidth{\itshape Acta Physica Polonica B Proceedings Supplement\hfil
      {\upshape\bfseries 20}{\upshape, ???? (2027)}}\vskip2pt\hrule\@height0.6pt}}}%
  \let\@evenhead\@oddhead
  \let\@mkboth\@gobbletwo\let\sectionmark\@gobble\let\subsectionmark\@gobble}
\g@addto@macro\ps@headings{\def\@oddfoot{\hfil\small(\ThePage)\hfil}\let\@evenfoot\@oddfoot}
\makeatother

\title{PROBING ANOMALOUS HIGGS COUPLINGS AT A FUTURE MUON--PROTON COLLIDER%
\thanks{Presented at the ``Ultra-cold muons for the LHC'' Workshop, Krak\'ow, Poland, September 24--26, 2026.}}
\author{Thapelo Leboho$^{a}$, Mukesh Kumar$^{a}$, Edward Nkadimeng$^{a}$, Pramod Sharma$^{a}$,
Bruce Mellado$^{a,b}$
\address{$^{a}$Institute for Collider Particle Physics, School of Physics,\\
University of the Witwatersrand, Johannesburg, South Africa\\[1mm]
$^{b}$Institute of High Energy Physics, Chinese Academy of Sciences, Beijing, China}
}
\headtitle{Probing anomalous Higgs couplings at a future muon--proton collider}
\headauthor{T. Leboho et al.}

\begin{document}
\maketitle
\begin{abstract}
We investigate the sensitivity of a future muon--proton collider to the Standard Model (SM) Higgs trilinear self-coupling and to anomalous $hhh$, $hWW$ and $hhWW$ interactions through charged-current di-Higgs production, $\mu^+p\to\bar\nu_\mu hhj$, with both Higgs bosons decaying to $b\bar b$. We consider a $500$~GeV $\mu^+$ beam colliding with the $7$~TeV proton beam ($\sqrt{s}\simeq 3.7$~TeV). We derive expected one-parameter 95\% confidence intervals using the fiducial rate and the signed azimuthal angle between the missing transverse momentum and the forward jet. At $1~\abinv$, the expected 95\% CL interval for the Higgs trilinear self-coupling modifier is $\gone{hhh}\in[-0.14,\,1.43]$, with all other couplings fixed to their SM values. At $10~\abinv$, the interval narrows to $[-0.20,\,0.34]$, and the momentum-dependent $hWW$ coupling is constrained to $\gtwo{hWW}\in[-0.009,\,0.007]$. At this higher luminosity, the signed-angle distribution provides sensitivity to the sign of the CP-odd couplings and reduces their interval widths by about 20\% relative to the rate-only fit. Requiring the reconstructed di-Higgs invariant mass to be below $1$~TeV weakens the constraints by at most 12\% at $1~\abinv$.
\end{abstract}

\section{Introduction}
\label{sec:intro}
The Higgs trilinear self-coupling probes the shape of the Higgs potential. Charged-current di-Higgs production is directly sensitive to this coupling and also receives contributions from the $hWW$ and $hhWW$ vertices. At lepton--hadron colliders, the forward jet and missing transverse momentum provide a characteristic signature of charged-current $WW$ fusion~\cite{LHeC2012,LHeC2021}. For the FCC-he, with a $60$~GeV electron beam and a $50$~TeV proton beam, Ref.~\cite{Kumar2017} studied the sensitivity to the Higgs self-coupling and anomalous $hWW$ and $hhWW$ interactions. The azimuthal angle between the missing transverse momentum and the forward jet provides information on the Lorentz structure of these interactions~\cite{Biswal2012,Kumar2017}.

Following Ref.~\cite{Kumar2017}, we parametrise the $hhh$ interaction through a rescaling of the Standard Model (SM) trilinear Higgs self-coupling and an additional derivative term. For the $hWW$ and
$hhWW$ vertices, we consider CP-even and CP-odd Lorentz structures. The corresponding interaction terms are
\begin{align}
\mathcal{L}^{(3)}_{hhh} &= \frac{m_h^2}{2v}\big(1-\gone{hhh}\big)h^3 + \frac{1}{2v}\,\gtwo{hhh}\,h\,\partial_\mu h\,\partial^\mu h,\\
\mathcal{L}^{(3)}_{hWW} &= -g\Big[\frac{\gone{hWW}}{2m_W}W^{\mu\nu}W^\dagger_{\mu\nu}h
   +\frac{\gtwo{hWW}}{m_W}\big(W^\nu\partial^\mu W^\dagger_{\mu\nu}h+\text{h.c.}\big)\nonumber\\
   &\qquad\;+\frac{\gt{hWW}}{2m_W}W^{\mu\nu}\widetilde W^\dagger_{\mu\nu}h\Big],\\
\mathcal{L}^{(4)}_{hhWW} &= -g^2\Big[\frac{\gone{hhWW}}{4m_W^2}W^{\mu\nu}W^\dagger_{\mu\nu}h^2
   +\frac{\gtwo{hhWW}}{2m_W^2}\big(W^\nu\partial^\mu W^\dagger_{\mu\nu}h^2+\text{h.c.}\big)\nonumber\\
   &\qquad\;+\frac{\gt{hhWW}}{4m_W^2}W^{\mu\nu}\widetilde W^\dagger_{\mu\nu}h^2\Big],
\end{align}
Here $W_{\mu\nu}=\partial_\mu W_\nu-\partial_\nu W_\mu$ and $\widetilde W_{\mu\nu}=\tfrac12\epsilon_{\mu\nu\rho\sigma}W^{\rho\sigma}$. The SM trilinear Higgs self-coupling corresponds to $\gone{hhh}=1$. The SM is recovered when all anomalous coefficients vanish. The derivative $g^{(1)}_{hWW,hhWW}$ and $g^{(2)}$ interactions are CP even, and the $\tilde g$ interactions are CP odd. We vary the eight selected coefficients one at a time, fixing the other parameters to their SM values.

Motivated by the muon--proton collider proposal of Ref.~\cite{Piotrzkowski:2026xld, Piotrzkowski:2026gjg}, we consider collisions between a $500$~GeV $\mu^+$ beam and the $7$~TeV Large Hadron Collider (LHC) proton beam, corresponding to $\sqrt{s}\simeq 3.7$~TeV. We use the fiducial rate and the azimuthal-angle distribution to constrain the selected interactions and assess how strongly the projected limits depend on events with large di-Higgs invariant masses.

\begin{figure}[t]\centering
\includegraphics[width=0.9\linewidth]{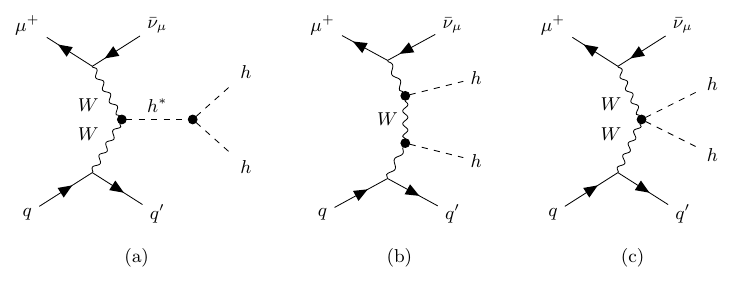}
\caption{Leading-order Feynman diagrams for the process $\mu^+p\to\bar\nu_\mu hhj$.}
\label{fig}
\end{figure}

Figure~\ref{fig} shows the leading-order Feynman diagrams for the process $\mu^+p\to\bar\nu_\mu hhj$. The amplitudes involving the $hhh$, $hWW$ and $hhWW$ vertices interfere, making both the production rate and the kinematic distributions sensitive to these interactions. We use the fiducial rate and the azimuthal-angle distribution to assess the sensitivity to variations in each coupling.

\section{Analysis and event selection}

We generate $\mu^+p\to\bar\nu_\mu hhj$ events with both Higgs bosons decaying to $b\bar b$ using {\tt{MadGraph5}}~\cite{MG5} and the interactions in Section~\ref{sec:intro}. To assess the sensitivity to each coupling, we vary it independently while fixing the remaining couplings to their
SM values. Consequently, for each one-parameter scan the cross section and the bin yields depend quadratically on the parameter, $\sigma(g)=c_0+c_1g+c_2g^2$. Here $g$ denotes the scan parameter. The SM signal cross section before tagging is $0.167$~fb. The backgrounds with four genuine $b$ quarks are $4b+j$, $hZj$, $hb\bar bj$ and $ZZj$, with $Z,h\to b\bar b$ where applicable. The $t\bar tj$ and $2b+$jets backgrounds enter through mistags.

The analysis is performed at parton level, with the energies of the $b$-quarks and jets smeared according to a 12\% resolution. We apply a $b$-tagging efficiency of 70\% to each $b$-quark and a mistag rate of 1\% to light jets. The event selection requires (i) at least four $b$-jets with $p_T>20$~GeV and $|\eta|<5$, (ii) a forward light jet $J$ with $\eta_J<-1.5$, along the proton direction, (iii) missing transverse energy $\MET>10$~GeV, (iv) $M_{4b}>200$~GeV, and (v) two $b\bar b$ pairs with invariant masses in the range $[105,145]$~GeV. The resulting event yields at $1~\abinv$ are presented in Table~\ref{tab:cutflow}. The Higgs-mass windows substantially suppress the $hZj$ and $ZZj$ backgrounds, as the $Z\to b\bar b$ invariant mass lies below the selected range. After all selection cuts, we obtain $S=22.3$ signal and $B=5.2$ background events, corresponding to a significance of $S/\sqrt{S+B}=4.25$. Table~\ref{tab:wp} presents the dependence of the yields on the $b$-tagging working point. Since the signal and dominant backgrounds scale with the same factor, $\epsilon_b^4$, their ratio $S/B$ is approximately unchanged, while the significance varies with the overall event yield. The effect of the mistag rate is negligible. The limits presented below are calculated using a 70\% $b$-tagging efficiency and a selection that reproduces the cut-flow yields to within 1\%.
\begin{table}[t]\centering\small
\caption{Expected events at $1~\abinv$ after each selection step for 70\% $b$-tagging and 1\% mistag rate. The signal corresponds to $166.7$ events before tagging ($\sigma=0.167$~fb). ``Mistag'' collects $t\bar tj$ (hadronic and semi-leptonic) and $2b+$jets, weighted by $\epsilon_b^2\epsilon_{\rm mis}^2=4.9\times10^{-5}$. $^{*}$After tagging weights. ``Total'' is the total background; the Higgs-mass windows are $[105,145]$~GeV.}
\label{tab:cutflow}
\footnotesize
\setlength{\tabcolsep}{2.6pt}
\begin{tabular}{@{}lrrrrrrrr@{}}
\toprule
Selection & Signal & $4b$ & $hZj$ & $hb\bar bj$ & $ZZj$ & Mistag & Total & $\frac{S}{\sqrt{S+B}}$\\
\midrule
Initial$^{*}$ & 40.02 & 130.50 & 128.67 & 96.88 & 139.57 & 39.02 & 534.6 & 1.67\\
$N_b\ge4$                    & 31.27 & 23.07  & 90.02  & 28.99 & 86.46  & 7.62  & 236.2 & 1.91\\
$\eta_J<-1.5$                & 25.95 & 2.99   & 75.34  & 14.24 & 59.50  & 0.45  & 152.5 & 1.94\\
$\MET>10$ GeV                & 25.19 & 2.98   & 73.67  & 13.84 & 59.18  & 0.44  & 150.1 & 1.90\\
$M_{4b}>200$ GeV             & 25.19 & 1.71   & 73.49  & 13.59 & 55.89  & 0.40  & 145.1 & 1.93\\
Higgs windows & 22.27 & 0.00   & 4.26   & 0.60  & 0.30   & 0.033 & 5.19  & 4.25\\
\bottomrule
\end{tabular}
\end{table}

\begin{table}[t]\centering\small
\caption{Final yields at $1~\abinv$ for different $b$-tagging working points.}
\label{tab:wp}
\begin{tabular}{lrrr}
\toprule
Working point & $S$ & $B$ & $S/\sqrt{S+B}$\\
\midrule
No tagging ($\epsilon_b=1$)       & 92.8 & 21.5 & 8.68\\
60\% $b$-tag, 1\% mistag          & 12.0 & 2.80 & 3.12\\
\textbf{70\% $b$-tag, 1\% mistag} & \textbf{22.3} & \textbf{5.18} & \textbf{4.25}\\
70\% $b$-tag, 2\% mistag          & 22.3 & 5.28 & 4.24\\
90\% $b$-tag, 1\% mistag          & 60.9 & 14.1 & 7.03\\
\bottomrule
\end{tabular}
\end{table}

\section{Observables and results}
We define the signed azimuthal separation as $\Delta\phi_s=\phi(\vec p_T^{\,\rm miss})-\phi(J)$, wrapped into $(-\pi,\pi]$, where the muon beam defines the $+z$ direction. The unsigned separation is $\Delta\phi=|\Delta\phi_s|$. For this ordering of the angles, the sign of $\sin\Delta\phi_s$ is the sign of $\hat z\cdot(\vec p_{T,J}\times\vec p_T^{\,\rm miss})$. The signed distribution is sensitive to terms linear in the CP-odd couplings, analogous to signed tagging-jet angles in weak-boson fusion~\cite{Plehn2002,Hankele2006}. The unsigned distribution also carries information on the Lorentz structure of the interactions~\cite{Biswal2012,Kumar2017}. We define the asymmetries
\begin{equation}
A=\frac{N(\Delta\phi>\pi/2)-N(\Delta\phi<\pi/2)}{N},\quad A_s=\frac{N(\Delta\phi_s>0)-N(\Delta\phi_s<0)}{N}.
\end{equation}
The expected limits are obtained from the binned $\Delta\phi_s$ yields after all selection cuts. The yield in each bin $i$ is scaled with the integrated luminosity and parametrised through its quadratic dependence on the anomalous coupling. We construct the test statistic
\begin{equation}
\begin{gathered}
\chi^2(g)=\vec d^{\,T}V^{-1}\vec d,\qquad d_i=N_i(g)-N_i^{\rm SM},\\
V=\mathrm{diag}\big(\vec N^{\rm SM}\big)+\delta^2\,\vec N^{\rm SM}\big(\vec N^{\rm SM}\big)^T,
\end{gathered}
\end{equation}
where $N_i^{\rm SM}$ includes both the signal and background contributions, and $\delta=5\%$ represents a normalisation uncertainty fully correlated across bins. We use $\chi^2=3.84$ as the nominal 95\% CL threshold for each one-parameter projection. For comparison, the rate-only limit is obtained using a single bin, while the 20-bin signed-angle fit retains both rate and shape information. For CP-even couplings, the signed-angle analysis reproduces the result obtained with the unsigned angle in 10 bins to within 1\%.

\begin{figure}[t]\centering
\includegraphics[width=0.7\linewidth]{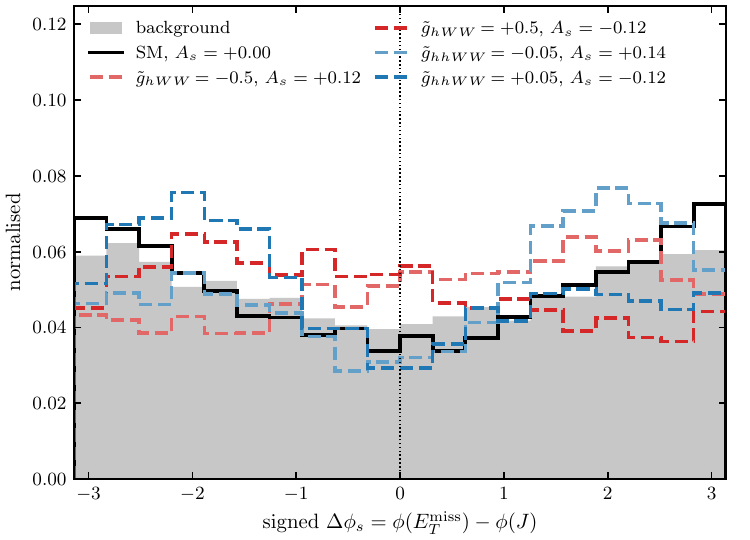}
\caption{Normalised distribution of the signed azimuthal angle $\Delta\phi_s$ after all cuts for the SM, the background and the benchmark samples. The CP-odd benchmarks $\gt{hWW}=\mp0.5$ (red) and $\gt{hhWW}=\mp0.05$ (blue) are shown dashed, with the lighter shade for the negative value; the two signs are mirror images of each other around $\Delta\phi_s=0$. The asymmetries in the legend are for the signal only.}
\label{fig:signed}
\end{figure}

After all selection cuts, the SM gives $A=+0.19$. The unsigned asymmetry responds differently to the two signs of the $g^{(2)}$ couplings. For example, $\gtwo{hWW}=\mp0.02$ gives $A=+0.03$ and $+0.17$, respectively. The signed asymmetry remains compatible with zero in these samples. In contrast, the fitted linear coefficient $c_1$ of the total rate is compatible with zero for the CP-odd benchmarks. Consequently, the rate constrains only $|\tilde g|$. However, these couplings generate a non-zero signed asymmetry of approximately equal magnitude and opposite sign for the two coupling signs, $A_s\simeq\pm0.11$, as illustrated in Fig.~\ref{fig:signed}. The SM signal and backgrounds yield $A_s\simeq0$. For the benchmarks shown, the projected sign discrimination is approximately $1\sigma$ for $1~\abinv$, increasing to $3.4\sigma$ at $10~\abinv$.
\begin{figure}[t]\centering
\includegraphics[
  width=\linewidth,
  height=0.52\textheight,
  keepaspectratio
]{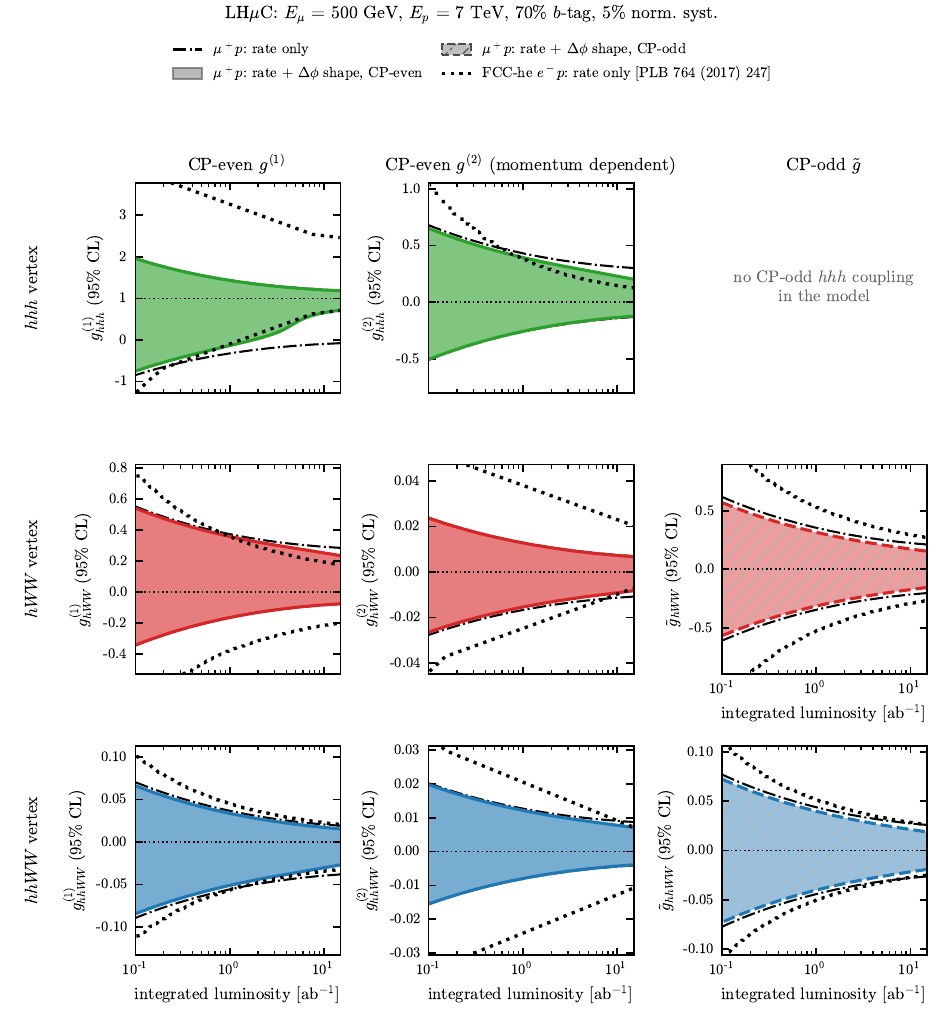}
\caption{Expected 95\% CL allowed regions as a function of the integrated luminosity. Rows: $hhh$, $hWW$ and $hhWW$ vertices (green, red and blue). In the first column, the existing $hhh$ panel uses the trilinear rescaling $1-\gone{hhh}$, while the gauge--Higgs panels use $g^{(1)}$. The second and third columns show $g^{(2)}$ and $\tilde g$, respectively. Dash-dotted lines: rate only; bands: rate and signed $\Delta\phi_s$ fit, with solid outline for CP-even and dashed, hatched for CP-odd couplings. Dotted lines: FCC-he, rate only~\cite{Kumar2017}. The range above about $1~\abinv$ is shown for comparison with the FCC-he.}
\label{fig:limits}
\end{figure}

Table~\ref{tab:limits} and Figure~\ref{fig:limits} show the projected 95\% CL intervals. We use $1~\abinv$ as the reference luminosity and also show higher-luminosity projections for comparison with the FCC-he study~\cite{Kumar2017}. Among the selected coefficients, the tightest numerical bounds are obtained for the momentum-dependent $hWW$ and $hhWW$ terms. The signed-angle fit yields the largest relative gain at $10~\abinv$ for $\gone{hhh}$, reducing its interval width by about 58\% compared with the rate-only fit. This gain reflects the weak rate discrimination between the SM point $\gone{hhh}=0$ and $\gone{hhh}=1$ in the scan. At the same luminosity, the signed-angle fit reduces the interval widths for the two CP-odd coefficients by approximately 20\%.

\begin{figure}[t]\centering
\includegraphics[width=0.65\textwidth]{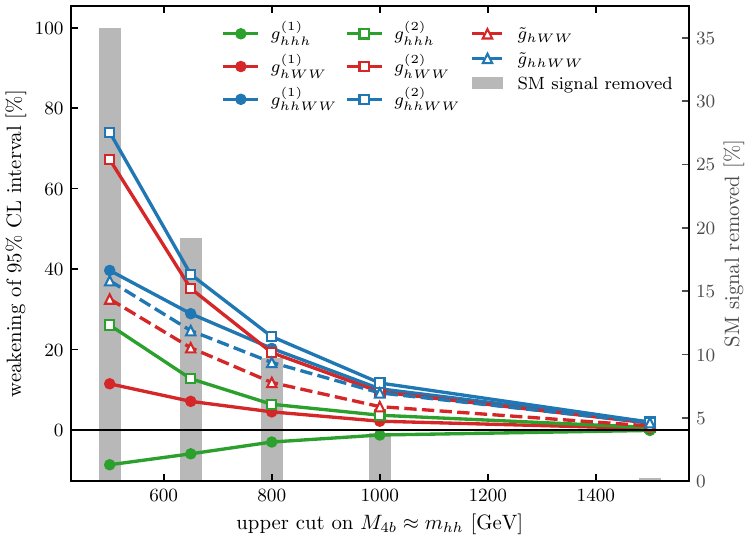}
\caption{Weakening of the expected 95\% CL intervals at $1~\abinv$ (rate and signed $\Delta\phi_s$) for all eight couplings as a function of an upper cut on $M_{4b}\approx m_{hh}$ (left axis), and the fraction of SM signal events removed by the cut (grey bars, right axis). Colours as in Fig.~\ref{fig:limits}; $g^{(1)}$ filled circles, $g^{(2)}$ open squares (solid lines), $\tilde g$ open triangles (dashed lines).}
\label{fig:eft}
\end{figure}

Finally, we test how strongly the projections depend on events with large di-Higgs invariant mass. Since the $g^{(2)}$ and $\tilde g$ vertices grow with momentum, these events can contribute substantially to the sensitivity of the corresponding couplings. We therefore repeat the analysis with an upper cut on $M_{4b}$, which approximates the di-Higgs invariant mass $m_{hh}$. Requiring $M_{4b}<1$~TeV removes only 4\% of the SM signal and weakens the limits at $1~\abinv$ by 2\% ($\gone{hWW}$), 10\% ($\gone{hhWW}$), 4\% ($\gtwo{hhh}$), 9\% ($\gtwo{hWW}$), 12\% ($\gtwo{hhWW}$), 6\% ($\gt{hWW}$) and 9\% ($\gt{hhWW}$), comparable to the findings for the FCC-he~\cite{Kumar2017}. The interval for $\gone{hhh}$ tightens by 2\% in this repeat fit. A more restrictive cut of $650$~GeV removes 19\% of the SM signal and weakens the limits on $\gtwo{hWW}$ and $\gtwo{hhWW}$ by 35--39\%, as shown for all eight couplings in Figure~\ref{fig:eft}.

\begin{table}[t]\centering\small
\caption{Expected one-parameter 95\% CL intervals from the rate-only fit and the signed-angle fit, which includes the rate.}
\label{tab:limits}
\footnotesize\setlength{\tabcolsep}{4pt}
\begin{tabular}{@{}lcccc@{}}
\toprule
 & \multicolumn{2}{c}{$1~\abinv$} & \multicolumn{2}{c}{$10~\abinv$}\\
\cmidrule(lr){2-3}\cmidrule(lr){4-5}
Coupling & rate & rate + shape & rate & rate + shape\\
\midrule
$\gone{hhh}$ & $[-0.44,\,1.32]$ & $[-0.43,\,1.14]$ & $[-0.21,\,1.09]$ & $[-0.20,\,0.34]$\\
$\gtwo{hhh}$  & $[-0.26,\,0.43]$ & $[-0.26,\,0.39]$ & $[-0.14,\,0.31]$ & $[-0.14,\,0.23]$\\
$\gone{hWW}$  & $[-0.17,\,0.37]$ & $[-0.17,\,0.36]$ & $[-0.08,\,0.29]$ & $[-0.08,\,0.25]$\\
$\gtwo{hWW}$  & $[-0.017,\,0.013]$ & $[-0.015,\,0.013]$ & $[-0.011,\,0.007]$ & $[-0.009,\,0.007]$\\
$\gt{hWW}$    & $[-0.35,\,0.35]$ & $[-0.32,\,0.32]$ & $[-0.22,\,0.22]$ & $[-0.17,\,0.17]$\\
$\gone{hhWW}$ & $[-0.056,\,0.037]$ & $[-0.051,\,0.034]$ & $[-0.039,\,0.021]$ & $[-0.030,\,0.017]$\\
$\gtwo{hhWW}$ & $[-0.008,\,0.013]$ & $[-0.008,\,0.012]$ & $[-0.004,\,0.009]$ & $[-0.004,\,0.008]$\\
$\gt{hhWW}$   & $[-0.044,\,0.044]$ & $[-0.040,\,0.040]$ & $[-0.028,\,0.027]$ & $[-0.022,\,0.021]$\\
\bottomrule
\end{tabular}
\end{table}

\section{Conclusion}
Charged-current di-Higgs production at a future muon--proton collider
provides sensitivity to the Higgs trilinear self-coupling and to anomalous
$hhh$, $hWW$ and $hhWW$ interactions. At $\sqrt{s}\simeq 3.7$~TeV and
$1~\abinv$, the expected one-parameter 95\% CL interval for the
trilinear deviation parameter is $\gone{hhh}\in[-0.43,\,1.14]$.
At $10~\abinv$, this interval narrows to $[-0.20,\,0.34]$, and the
momentum-dependent $hWW$ and $hhWW$ coefficients are probed at the
$10^{-2}$ level.

The signed azimuthal distribution between the missing transverse momentum
and the forward jet adds information on the Lorentz structure of the
interactions and provides sensitivity to the sign of the CP-odd
coefficients. At $1~\abinv$, requiring $M_{4b}<1$~TeV weakens the
projected constraints by at most 12\%, indicating limited sensitivity to
the removal of this high-mass region. A detector-level study incorporating
realistic reconstruction and multivariate selection is in preparation.

\bibliographystyle{unsrt}
\bibliography{biblio}
\end{document}